\documentclass[a4paper,fleqn]{cas-sc}

\usepackage[authoryear,longnamesfirst]{natbib}
\usepackage{amsmath,amssymb}
\usepackage{booktabs}
\usepackage{orcidlink}
\usepackage{xcolor}
\usepackage[normalem]{ulem}

\begin{document}

\let\WriteBookmarks\relax
\def\floatpagepagefraction{1}
\def\textpagefraction{.001}

\shorttitle{Inertial-Range Suppression in Sub-Alfv\'{e}nic PSBL Turbulence}
\shortauthors{Mani~K~Chettri et~al.}

\title[mode=title]{Inertial-Range Suppression and Ponderomotive Density Cavitation in Broadband Sub-Alfv\'{e}nic Turbulence under Plasma Sheet Boundary Layer Conditions}

\author[1]{Mani K Chettri}[orcid=0009-0000-1368-9263]
\fnmark[1]
\credit{Conceptualization, Methodology, Software, Formal Analysis, Writing -- Original Draft}

\author[1]{Vivek Shrivastav}[orcid=0009-0003-6908-7263]
\credit{Formal Analysis, Writing -- Review \& Editing}

\author[1]{Britan Singh}[orcid=0009-0009-9813-3402]
\credit{Formal Analysis, Writing -- Review \& Editing}

\author[1]{Rupak Mukherjee}[orcid=0000-0003-3955-7116]
\credit{Supervision, Validation, Writing -- Review \& Editing}

\author[2]{Hemam D. Singh}[orcid=0009-0003-3061-8944]
\cormark[1]
\ead{hemam.singh@nsut.ac.in}
\credit{Supervision, Methodology, Validation, Writing -- Review \& Editing}

\affiliation[1]{organization={Department of Physics, Sikkim University},
            city={Gangtok}, postcode={737102}, state={Sikkim}, country={India}}

\affiliation[2]{organization={Department of Physics, Netaji Subhas University of Technology},
            city={New Delhi}, postcode={110078}, country={India}}

\cortext[1]{Corresponding author}
\fntext[1]{Email: mkchettri8@gmail.com}

\begin{abstract}
Kinetic Alfv\'{e}n waves (KAWs) are among the most pervasive electromagnetic fluctuations in magnetized astrophysical plasmas, from Earth's magnetospheric boundary layers to the turbulent intracluster medium of galaxy clusters. Their ponderomotive coupling to compressive density fluctuations remains incompletely understood in the broadband turbulent regime. We present two-dimensional pseudospectral simulations of the modified nonlinear Schr\"{o}dinger--magnetosonic (MNLS--MS) system governing KAW envelopes, initialized with a broadband power-law spectrum ($|\psi(\mathbf{k})|^2\propto k^{-5/6}$) spanning many interacting modes, at $\beta \sim 0.1$--$0.3$ representative of plasma sheet boundary layer (PSBL) conditions. A fourth-order Runge--Kutta scheme on a $256\times 256$ grid integrates the system to $t = 40$ (normalized), with total energy conserved to within $0.085\%$ in the undamped run; a damped run with dissipation loses $\sim 4\%$ of the magnetic energy over the same interval. The nonlinearity parameter $\chi_\mathrm{NL} \approx 0.25$ confirms broadband sub-Alfv\'{e}nic turbulence throughout. Magnetic field intensity and plasma density develop spatially intermittent, filamentary structures within the first few wave periods, consistent with ponderomotive density cavitation and plasma expulsion from wave-intense regions. The magnetic energy spectra show inertial-range suppression, with a rapid transition from injection ($k < 0.3$) to dissipation without an extended power-law cascade, in agreement with the moderate magnetic Reynolds number ($\mathrm{R_m} \sim 250$--$370$) of the simulation and the observationally constrained range for PSBL turbulence. These results provide numerical evidence that broadband KAW turbulence self-organizes into coherent density structures at kinetic scales, and that the spectral character of such turbulence is governed primarily by moderate-Reynolds-number constraints rather than by the wave physics alone.
\end{abstract}

\begin{keywords}
kinetic Alfv\'{e}n waves \sep modified nonlinear Schr\"{o}dinger equation \sep plasma turbulence \sep ponderomotive force \sep density cavities \sep pseudospectral method \sep plasma sheet boundary layer \sep astrophysical plasmas
\end{keywords}

\maketitle

\section{Introduction}
\label{sec:intro}

Alfv\'{e}nic turbulence is a universal feature of magnetized astrophysical plasmas, mediating the transfer of energy from the large fluid scales at which it is injected down to the kinetic scales of individual particles, where it is ultimately thermalized. At perpendicular spatial scales approaching the ion gyroradius $\rho_i$ or the ion inertial length $d_i$, ideal shear Alfv\'{e}n waves acquire dispersive corrections and transform into kinetic Alfv\'{e}n waves (KAWs), a transition that is observationally well established in the solar wind
\citep{leamon1999, bale2005, chen2013, sahraoui2009}, in Earth's magnetosphere
\citep{chaston2008, wygant2002}, and in the magnetosheath \citep{roberts2018, gershman2017}.
KAWs carry a finite parallel electric field $E_\parallel$ that opens a direct channel for wave-particle energy exchange through Landau damping \citep{howes2008, chen2019}, and at finite amplitude they exert a ponderomotive force on the bulk plasma, a low-frequency, gradient-driven pressure $\propto{-}\nabla|\delta B_y|^2/8\pi$, which expels ions from regions of high wave intensity and forms localized density depletions. These two effects, particle heating and density structuring, make KAWs important agents not only in Earth's magnetotail
\citep{wygant2002, stawarz2017, chettri2025mms} but also in the broader context of astrophysical
plasma heating. In solar coronal loops, parametric decay of standing Alfv\'{e}n waves generates compressive daughter modes with observationally testable signatures \citep{terradas2022}. In the inner heliosphere, coupled KAW and ion-acoustic wave dynamics have been studied using Parker Solar Probe data \citep{chettri2024raa}. In the interstellar medium, kinetic-scale Alfv\'{e}nic turbulence shapes the scattering properties of pulsars and governs cosmic-ray transport \citep{lazarian2006, yan2004}. In the turbulent intracluster medium of galaxy clusters, where magnetic fields thread hot, weakly collisional gas, analogous wave-plasma interactions on scales below the Coulomb mean free path have been invoked to explain observed magnetic field tangling and sub-Larmor density fluctuations \citep{brunetti2007, schekochihin2009}.

The theoretical framework for KAW nonlinear dynamics rests on foundational studies spanning several decades. \citet{hasegawa1976parametric} identified parametric decay as the primary instability by which a monochromatic KAW transfers energy to compressive daughter modes.
\citet{shukla1999} derived the coupled KAW--magnetosonic equations that describe ponderomotive
density cavity formation in the quasi-static limit and demonstrated the linear anti-correlation $n/n_0 \propto -|\delta B_y|^2$. More recently, three-dimensional hybrid simulations of circularly polarized Alfv\'{e}n waves by \citet{verscharen2024} have confirmed that the parametric decay instability operates across a broad spectrum of wave vectors. \citet{goldreich1995} and \citet{boldyrev2012} established the theoretical spectral scalings expected when Alfv\'{e}nic turbulence reaches a critically balanced cascade, providing benchmarks against which kinetic-scale simulations can be compared. Our group recently derived and validated a modified nonlinear Schr\"{o}dinger equation (MNLSE) framework that self-consistently incorporates KAW dispersion, ponderomotive nonlinearity, and phenomenological Landau damping, and tested it against MMS magnetosheath observations \citep{chettri2025damped}. Separately,
\citet{chettri2025mms} reported direct MMS observations of KAW turbulence signatures in the
outer PSBL, establishing the observational context for the present numerical study. Nevertheless, the ponderomotive scaling relation has been tested chiefly in analytical or single-wave-packet settings; how it behaves within a broadband fluctuation environment where multiple interacting modes compete and interfere remains much less well characterized.

This gap motivates the present work. We solve the coupled MNLS--magnetosonic system of
\citet{chettri2025damped} in two spatial dimensions using a pseudospectral method on a
$256\times 256$ grid, initialized with a broadband turbulent spectrum representative of PSBL conditions. The PSBL provides a well-observed astrophysical boundary layer where magnetic energy, particle fluxes, and broadband electromagnetic turbulence coexist at directly measurable intensities, giving concrete observational anchors for the simulation parameters. Our goals are fourfold: (1)~to examine how KAW ponderomotive coupling operates when the driving is turbulent rather than monochromatic; (2)~to characterize the spatial structure and temporal persistence of the resulting density fluctuations; (3)~to assess the spectral character of the magnetic energy cascade at moderate Reynolds numbers; and (4)~to provide diagnostics (energy conservation, nonlinearity parameter, and undamped-vs-damped energy comparison) that allow meaningful comparison with in-situ observations and theoretical expectations.

Section~\ref{sec:theory} presents the governing MNLS--MS equations and the ponderomotive scaling prediction. The numerical method, parameters, and initialization are described in Section~\ref{sec:method}. Results are presented in Section~\ref{sec:results}, followed by discussion in Section~\ref{sec:discussion} and a summary in Section~\ref{sec:conclusions}.

\section{Theoretical Framework}
\label{sec:theory}

We consider a magnetized plasma with a uniform ambient magnetic field $\mathbf{B}_0 = B_0\hat{z}$ and background density $n_0$. The total density is $N = n_0 + n$, with $n$ the perturbation. The KAW magnetic field perturbation is written as a slowly-varying complex envelope, $\delta B_y = \mathrm{Re}[\psi\,e^{i(k_0 x - \omega_0 t)}]$, where $\psi(x,z,t)$ evolves on timescales much longer than $\omega_0^{-1}$. In the slowly-varying envelope approximation, the time-averaged magnetic field intensity is $\langle|\delta B_y|^2\rangle = |\psi|^2/2$; we refer to $|\delta B_y|^2$ (as labelled in the figures) and $|\psi|^2$ interchangeably throughout, noting that they carry the same spatial and temporal structure.

\subsection{Modified Nonlinear Schr\"{o}dinger Equation}
\label{sec:theory:mnls}

Following \citet{chettri2025damped}, the slowly-varying envelope $\psi$ satisfies the modified nonlinear Schr\"{o}dinger (MNLS) equation
\begin{equation}
  i\frac{\partial\psi}{\partial t}
  + P\,\frac{\partial^2\psi}{\partial x^2}
  + Q\,|\psi|^2\,\psi
  + i\Gamma\,\psi = 0,
  \label{eq:mnls}
\end{equation}
whose three coefficients encode the dispersive, ponderomotive, and dissipative physics of the KAW, respectively.

The \textit{dispersion coefficient} $P$ follows from the KAW dispersion relation $\omega^2 = k_z^2 v_A^2(1 + k_\perp^2\rho_s^2)$:
\begin{equation}
  P = \frac{1}{2}\frac{\partial^2\omega}{\partial k_\perp^2}
    = \frac{v_A^2\rho_s^2 k_z^2}{2\,\omega_0(1+k_0^2\rho_s^2)},
  \label{eq:P}
\end{equation}
where $v_A = B_0/\sqrt{4\pi n_0 m_i}$ is the Alfv\'{e}n speed, $\rho_s = C_s/\Omega_i$ is the ion-sound gyroradius, $C_s = \sqrt{k_B T_e/m_i}$ is the ion-sound speed, and $\Omega_i = eB_0/(m_i c)$ is the ion gyrofrequency. The $\rho_s^2$ factor makes dispersion a purely kinetic effect, absent in ideal MHD.

The \textit{ponderomotive nonlinearity coefficient} $Q$ arises from the self-consistent coupling between wave pressure $|\psi|^2/8\pi$ and the magnetosonic density response:
\begin{equation}
  Q = \frac{\omega_0\,\Gamma_2}{C_s^2},
  \qquad
  \Gamma_2 = \frac{v_A^2}{4B_0^2}.
  \label{eq:Q}
\end{equation}
Because $Q > 0$, regions of enhanced $|\psi|^2$ expel plasma and decrease the local refractive index, reinforcing the growth of filamentary intensity peaks through ponderomotive self-focusing
\citep{shukla1999, hasegawa1976parametric}.

The \textit{Landau damping coefficient} $\Gamma$ approximates the collisionless electron (and ion) damping of the KAW envelope. Following \citet{chettri2025damped}, it is evaluated from the imaginary part of the linearized kinetic dispersion function at $(k_0,\omega_0)$:

\begin{equation}
  \Gamma = |\,\mathrm{Im}[\omega_\mathrm{kin}(k_0)]|
  \label{eq:Gamma}
\end{equation}
In the low-$\beta$ PSBL regime ($\beta \sim 0.1$--$0.3$, $T_i/T_e = 2$), resonant damping is well below the wave frequency, so $\Gamma$ acts as a perturbative correction to the dispersive and nonlinear dynamics. It is retained for consistency with the high-$\beta$ magnetosheath analysis of \citet{chettri2025damped}; the full kinetic treatment of velocity-space resonances is deferred to gyrokinetic approaches such as those described in \citet{schekochihin2009} and \citet{howes2008}.

\subsection{Ponderomotively Driven Density Equation}
\label{sec:theory:density}

The ponderomotive force $\mathbf{f}_\mathrm{pond} \propto -\nabla|\psi|^2$ drives compressive fluctuations through the magnetosonic branch. Following \citet{shukla1999} and
\citet{chettri2025damped}, the general form of the density equation is
\begin{equation}
  \left(\frac{\partial^2}{\partial t^2}
        - C_s^2\,\nabla^2\right)\frac{n}{n_0}
  = \Gamma_2\,\nabla^2|\psi|^2.
  \label{eq:msw_general}
\end{equation}
In the strongly magnetized PSBL regime ($\beta \ll 1$), compressive perturbations propagate preferentially along the ambient field $\mathbf{B}_0 = B_0\hat{z}$, and cross-field acoustic transport is suppressed by the strong magnetic pressure. The density response is therefore governed primarily by the field-aligned component of the ponderomotive gradient. We accordingly adopt the parallel-propagation approximation $\nabla^2 \to \partial^2/\partial z^2$, reducing the density equation to
\begin{equation}
  \left(\frac{\partial^2}{\partial t^2}
        - C_s^2\,\frac{\partial^2}{\partial z^2}\right)\frac{n}{n_0}
  = \Gamma_2\,\frac{\partial^2}{\partial z^2}|\psi|^2.
  \label{eq:msw}
\end{equation}

The strong ambient field $\mathbf{B}_0$ resists perpendicular plasma displacement, so ponderomotively driven density perturbations propagate preferentially along the field. At the plasma parameters of Table~\ref{tab:params}, $C_s/v_A \approx 0.31$, so the omitted perpendicular term $C_s^2\,\partial^2/\partial x^2$ is suppressed relative to the retained parallel term by a factor $C_s^2/v_A^2 \approx 0.09$; the approximation introduces corrections of order $9\%$, within the first-order accuracy of the MNLS--MS framework and consistent with the canonical treatment of \citet{shukla1999}. The simulations also operate in the quasi-static limit ($\chi_\mathrm{NL}\approx 0.25 \ll 1$); in this limit the common spatial operator cancels from both sides of equation~(\ref{eq:msw_general}), and the ponderomotive scaling relation~(\ref{eq:scaling}) is independent of whether $\partial^2/\partial z^2$ or $\nabla^2$ is used. The central anti-correlation results of Section~\ref{sec:results:density} are therefore unaffected by this approximation. Inclusion of the full Laplacian is identified as a direction for future work in Section~\ref{sec:discussion:limits}. Equations~(\ref{eq:mnls}) and~(\ref{eq:msw}) form the coupled MNLS--MS system solved in the present simulations.

In the quasi-static limit ($|\partial^2/\partial t^2| \ll C_s^2 k_z^2$), the common parallel Laplacian cancels and equation~(\ref{eq:msw}) reduces to the ponderomotive scaling relation
\begin{equation}
  \frac{n}{n_0}
  \approx -\frac{\Gamma_2}{C_s^2}\,|\psi|^2
          = -\frac{v_A^2}{4B_0^2 C_s^2}\,|\psi|^2,
  \label{eq:scaling}
\end{equation}
predicting a linear anti-correlation between local wave intensity and plasma density with slope $-\Gamma_2/C_s^2$.

The density perturbations generated by equation~(\ref{eq:msw}) represent quasi-modes continuously maintained by the KAW ponderomotive source, rather than freely propagating ion acoustic waves. In the magnetotail, where $T_i/T_e \sim 1$--$3$, ion Landau damping strongly suppresses free acoustic propagation \citep{verscharen2024, terradas2022}. Energy transferred from the KAW envelope into these density quasi-modes can thermalize rapidly via ion Landau damping, providing a direct ion heating channel not resolved by the present fluid model but important for the astrophysical interpretation.

We note that the cubic nonlinearity $Q|\psi|^2\psi$ in equation~(\ref{eq:mnls}) already embeds the quasi-static density response (equation~\ref{eq:scaling}) within the envelope evolution: the coefficient $Q = \omega_0\Gamma_2/C_s^2$ is precisely the ponderomotive coupling strength evaluated in the adiabatic limit. The dynamical density equation~(\ref{eq:msw}) is solved separately to resolve finite-frequency corrections to this adiabatic balance and to provide the spatially resolved density field for diagnostic comparison with spacecraft observations. Feedback of the dynamical density on the envelope beyond the quasi-static limit is a higher-order correction, small by $\mathcal{O}(\chi_\mathrm{NL}^2)$ in the sub-Alfv\'{e}nic regime explored here ($\chi_\mathrm{NL}\approx 0.25$).

\section{Numerical Method and Simulation Parameters}
\label{sec:method}

\subsection{Algorithm}

The MNLS--MS system, equations~(\ref{eq:mnls}) and~(\ref{eq:msw}), is solved in two spatial dimensions ($x$, $z$) using a pseudospectral method following the approach of
\citet{chettri2025damped}. All spatial derivatives are computed in Fourier space via the Fast
Fourier Transform (FFT), which provides spectral accuracy for smooth fields. Time integration uses a classical fourth-order Runge--Kutta (RK4) scheme with time step $\Delta t = 10^{-5}$ (normalized).

At each time step, the complex envelope $\psi$ and density field $n/n_0$ are transformed to wavenumber space. Spatial derivative operators are applied algebraically, and the $2/3$-rule dealiasing mask is applied to suppress aliasing from the nonlinear terms $|\psi|^2\psi$ and $|\psi|^2$ before transforming back to physical space. The Landau damping term $\Gamma\psi$ is also evaluated spectrally. Grid-scale noise is controlled by a high-order hyperviscosity operator $-\nu_h k^6$ applied to wavenumbers above $3k_\mathrm{max}/4$, with coefficient $\nu_h = 10^{-9}$ chosen to confine energy dissipation to the grid scale without contaminating physical-scale dynamics. The density equation~(\ref{eq:msw}) is cast as a first-order system by introducing the auxiliary variable $w = \partial(n/n_0)/\partial t$, and both $n/n_0$ and $w$ are advanced together with the envelope by RK4. The parallel-derivative operator $\partial^2/\partial z^2$ in equation~(\ref{eq:msw}) is evaluated as $-k_z^2$ in Fourier space.

\subsection{Simulation Parameters}

Table~\ref{tab:params} lists all simulation parameters. Plasma parameters are chosen to represent the near-Earth PSBL at $X_\mathrm{GSM}\approx -7\,R_E$, based on observational compilations from \citet{stawarz2017} and statistical analyses in \citet{du2011} and
\citet{zhou2012}. The values $B_0 = 63$~nT, $n_0 = 0.72$~cm$^{-3}$, and $T_e = 2.6$~keV are
taken directly from observations, and the temperature ratio $T_i/T_e = 2$ is consistent with Hydra spacecraft measurements in this region \citep{shrivastava2015}. Together these yield $\beta\sim 0.1$--$0.3$, placing the simulation in the warm plasma regime where electromagnetic KAW generation and propagation are most efficient \citep{denton2010}.

Spatial coordinates are normalized to the ion inertial length $d_i = c/\omega_{pi}$, time to $\Omega_i^{-1}$, magnetic field to $B_0$, and density to $n_0$. In these normalized units, $v_A$ and $\rho_s$ are both set to unity for computational convenience; the physical plasma beta enters through the dimensionless coefficients $P$, $Q$, and $\Gamma$, which are computed from the PSBL parameters listed in Table~\ref{tab:params}.

\begin{table}[width=\linewidth,cols=2,pos=h!]
\caption{Simulation parameters. Observed plasma values $B_0$, $n_0$, and $T_e$ are from
\citet{stawarz2017}; $T_i/T_e$ from \citet{shrivastava2015}.}
\label{tab:params}
\renewcommand{\arraystretch}{1.2}
\begin{tabular*}{\tblwidth}{@{} LL @{}}
\toprule
Parameter & Value \\
\midrule
Grid resolution               & $256\times 256$                  \\
Domain size                   & $L_x\times L_z = (2\pi/0.1)^2$  \\
Time step $\Delta t$          & $10^{-5}$                        \\
$t_\mathrm{max}$              & 40                               \\
Alfv\'{e}n speed $v_A$        & 1.0                              \\
Ion-sound gyroradius $\rho_s$ & 1.0                              \\
Plasma beta $\beta$           & 0.1--0.3                         \\
Hyperviscosity $\nu_h$        & $10^{-9}$                        \\
$B_0$                         & 63~nT                            \\
$n_0$                         & 0.72~cm$^{-3}$                   \\
$T_e$                         & 2.6~keV                          \\
$T_i/T_e$                     & 2                                \\
\bottomrule
\end{tabular*}
\end{table}

\subsection{Initialization}

The simulation begins with a broadband turbulent spectrum, in deliberate contrast to single-wave-packet approaches. The initial complex envelope in Fourier space is assigned a power-law amplitude $|\psi(\mathbf{k})|^2 \propto (k/k_\mathrm{min})^{-5/6}$ over the injection range $k_\mathrm{min} = 0.08$ to $k_\mathrm{max} = 0.5$ (normalized), with random phases drawn uniformly from $[0, 2\pi)$ and a fixed seed for reproducibility. The real-space field is normalized to a standard deviation equal to the desired amplitude $A_B = 0.5$. The density is initialized with a correlated but phase-independent field at amplitude $0.4\,A_B$, and the auxiliary velocity variable $w$ is set from approximate acoustic balance. This procedure places turbulent energy across many modes from the outset, eliminating the artificial transient that arises when a single wave packet is allowed to self-organize under ponderomotive forcing.

\section{Simulation Results}
\label{sec:results}

\subsection{Energy Conservation and Nonlinearity}
\label{sec:results:energy}

\begin{figure}
  \centering

  \includegraphics[width=0.8\textwidth]{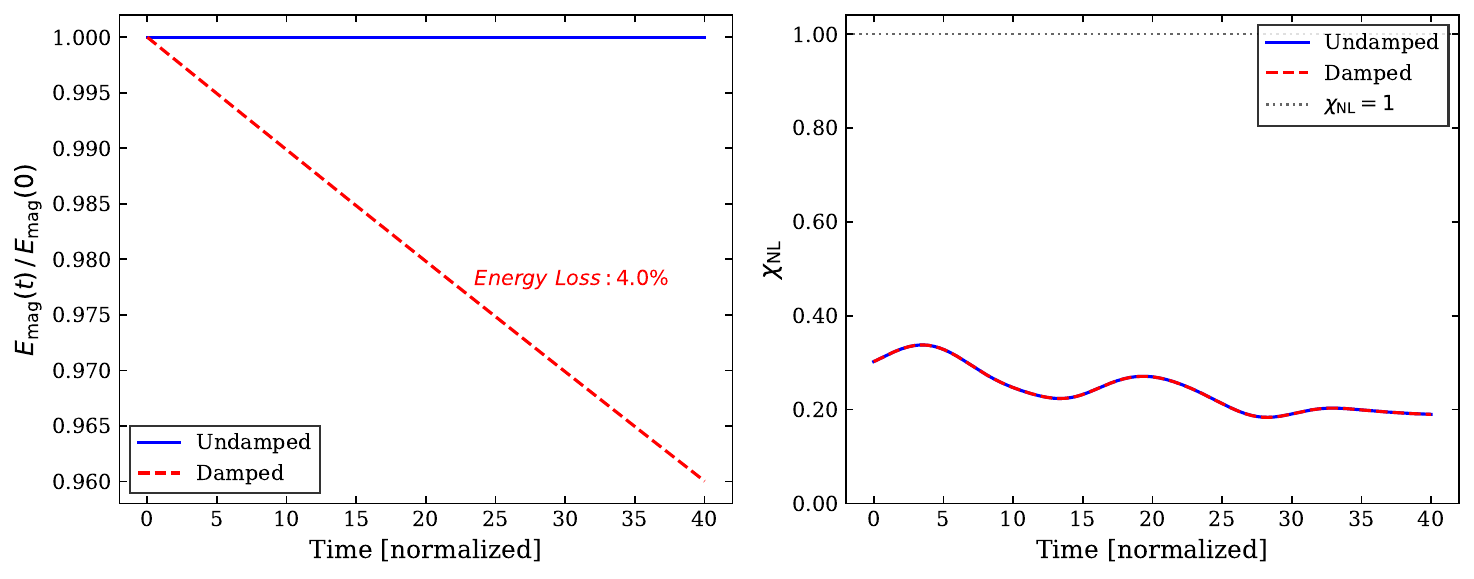}
  \caption{ (a)~Normalized magnetic energy $E_\mathrm{mag}(t)/E_\mathrm{mag}(0)$: the undamped run
    (blue solid, $\Gamma = 0$) drifts only $-0.085\%$, while the damped run (red dashed) loses $\sim 4\%$ to Landau dissipation, confirming that hyperviscosity and Landau damping act at their expected levels. (b)~Nonlinearity parameter $\chi_\mathrm{NL}$ as a function of time for both runs; values near $0.25$, well below the strong-turbulence threshold $\chi_\mathrm{NL} = 1$ (grey dotted line), confirm sub-Alfv\'{e}nic operation regardless of damping.}%
  \label{fig:validation}
\end{figure}

Figure~\ref{fig:validation} shows two diagnostic quantities tracked throughout the run. Panel~(a) displays the normalized magnetic energy $E_\mathrm{mag}(t)/E_\mathrm{mag}(0)$ for two cases: an undamped run ($\Gamma = 0$, blue solid) and a damped run with the Landau coefficient evaluated from equation~(\ref{eq:Gamma}) (red dashed). In the undamped case the energy is essentially flat, drifting by only $-0.085\%$ over the full integration to $t = 40$; this drift is attributable entirely to the hyperviscosity $-\nu_h k^6$, acting exclusively above $3k_\mathrm{max}/4$, well above the injection scales where most energy resides ($k \lesssim 0.3$). In the damped case, Landau dissipation removes $\sim 4\%$ of the magnetic energy by $t = 40$, confirming that the Landau channel produces a measurable but modest energy drain at these parameters. Because the scalar $\Gamma$ is evaluated at a single carrier wavenumber and cannot consistently represent damping across the full broadband spectrum ($k = 0.08$--$0.5$), where the kinetic rate $\gamma(k)$ varies by roughly an order of magnitude, we retain $\Gamma = 0$ to verify energy conservation, providing a validation benchmark for the numerical code.

Panel~(b) shows the nonlinearity parameter $\chi_\mathrm{NL}$ for both the undamped and damped runs, calculated as the ratio of the standard deviation of the product $|\delta B_y|\,|n/n_0|$ to that of $|\delta B_y|$ alone, evaluated over the simulation domain. The two curves are virtually identical, confirming that the sub-Alfv\'{e}nic regime is robust to the inclusion of Landau damping. The values remain in the range $0.2$--$0.28$ throughout, well below the critical threshold $\chi_\mathrm{NL} = 1$ at which the transition to strong turbulence occurs. The simulation therefore operates in the sub-Alfv\'{e}nic, moderately nonlinear regime characteristic of quieter PSBL intervals and of many other astrophysical boundary layers \citep{galtier2000, matthaeus2011}.

\subsection{Temporal Evolution of Magnetic Structures}
\label{sec:results:magnetic}

\begin{figure}
  \centering
  \includegraphics[width=0.9\textwidth]{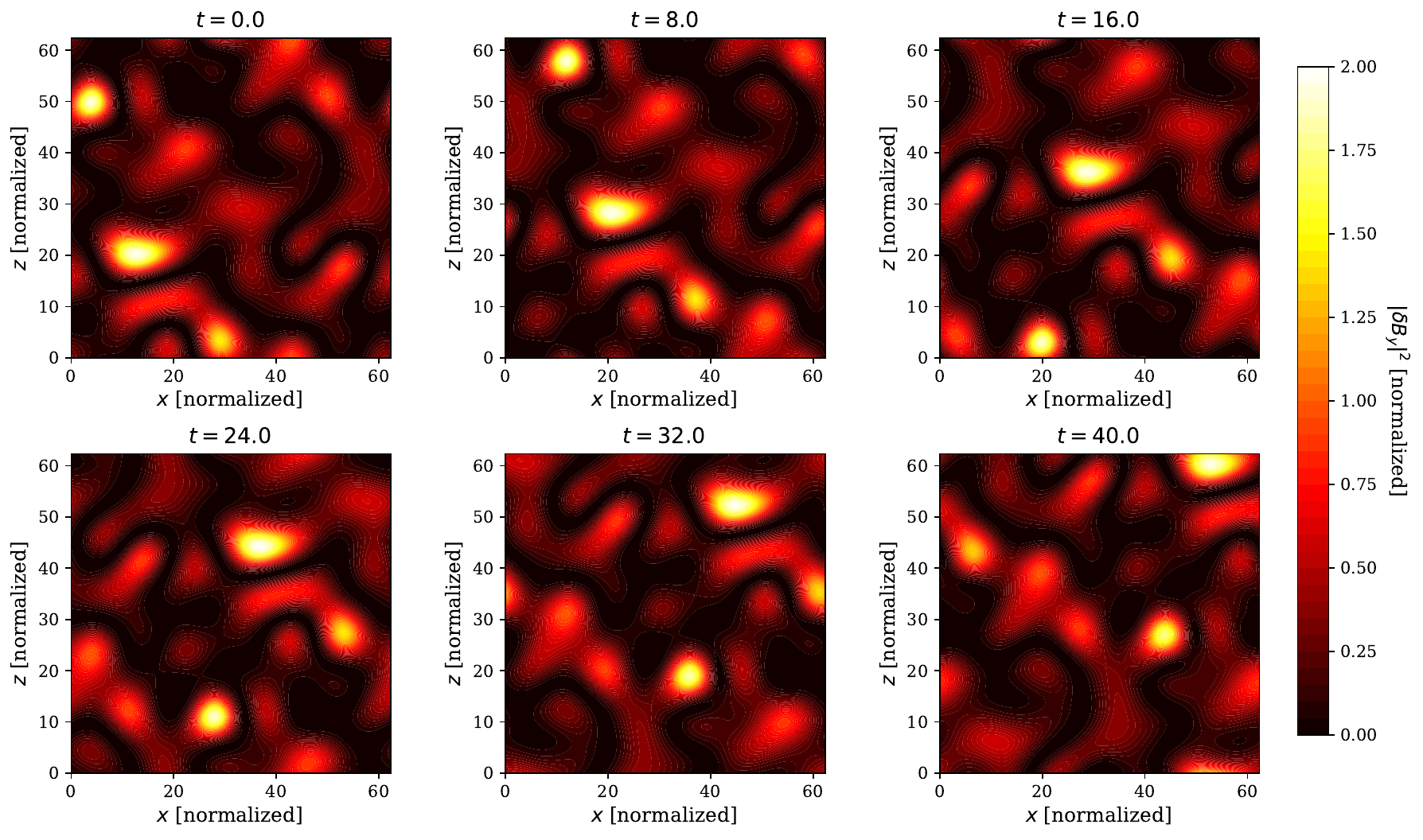}
  \caption{Temporal evolution of the magnetic field intensity $|\delta B_y|^2$ (normalized) at
  $t = 0, 8, 16, 24, 32$ and $ 40$. Elongated, filamentary structures emerge from the initial broadband spectrum and persist throughout. Peak intensities reach $\sim 1.6$ in isolated regions while the background stays near zero, indicating notable spatial intermittency characteristic of kinetic-scale turbulence.}
  \label{fig:magnetic}
\end{figure}

Figure~\ref{fig:magnetic} shows the two-dimensional distribution of the magnetic field intensity $|\delta B_y|^2$ at six times from $t = 0$ to $t = 40$. The initialization at $t = 0$ is spatially irregular without distinct preferred structures. By $t = 8$, elongated filaments of elevated intensity have already emerged from the background, indicating that the broadband turbulence self-organizes on a timescale of order several wave periods, broadly consistent with the nonlinear interaction time $\tau_\mathrm{nl} \sim (Q|\delta B_y|^2)^{-1}$ at these amplitudes. By $t = 16$ the filaments are more pronounced, with peak normalized intensities exceeding $1.6$ in isolated regions while the surrounding plasma carries much lower intensity.

Through $t = 24$, $32$, and $40$ the filamentary pattern continues to evolve: structures shift, merge, and re-form. The general character, a spatially intermittent field with localized intensity peaks surrounded by a quiescent background, persists throughout. Such intermittency is characteristic of kinetic-scale turbulence and is consistent with observations of magnetic field structures in the PSBL and magnetosheath \citep{gershman2017, chen2019, sahraoui2009, chettri2025mms}. The perpendicular extents of the structures, of order $10$--$20\,\rho_s$, are comparable to the scales of magnetic holes and enhancements reported by MMS in the plasma sheet boundary layer \citep{huang2017, gershman2017b}.

\subsection{Density Cavity Formation and Persistence}
\label{sec:results:density}

\begin{figure}
  \centering
  \includegraphics[width=0.9\textwidth]{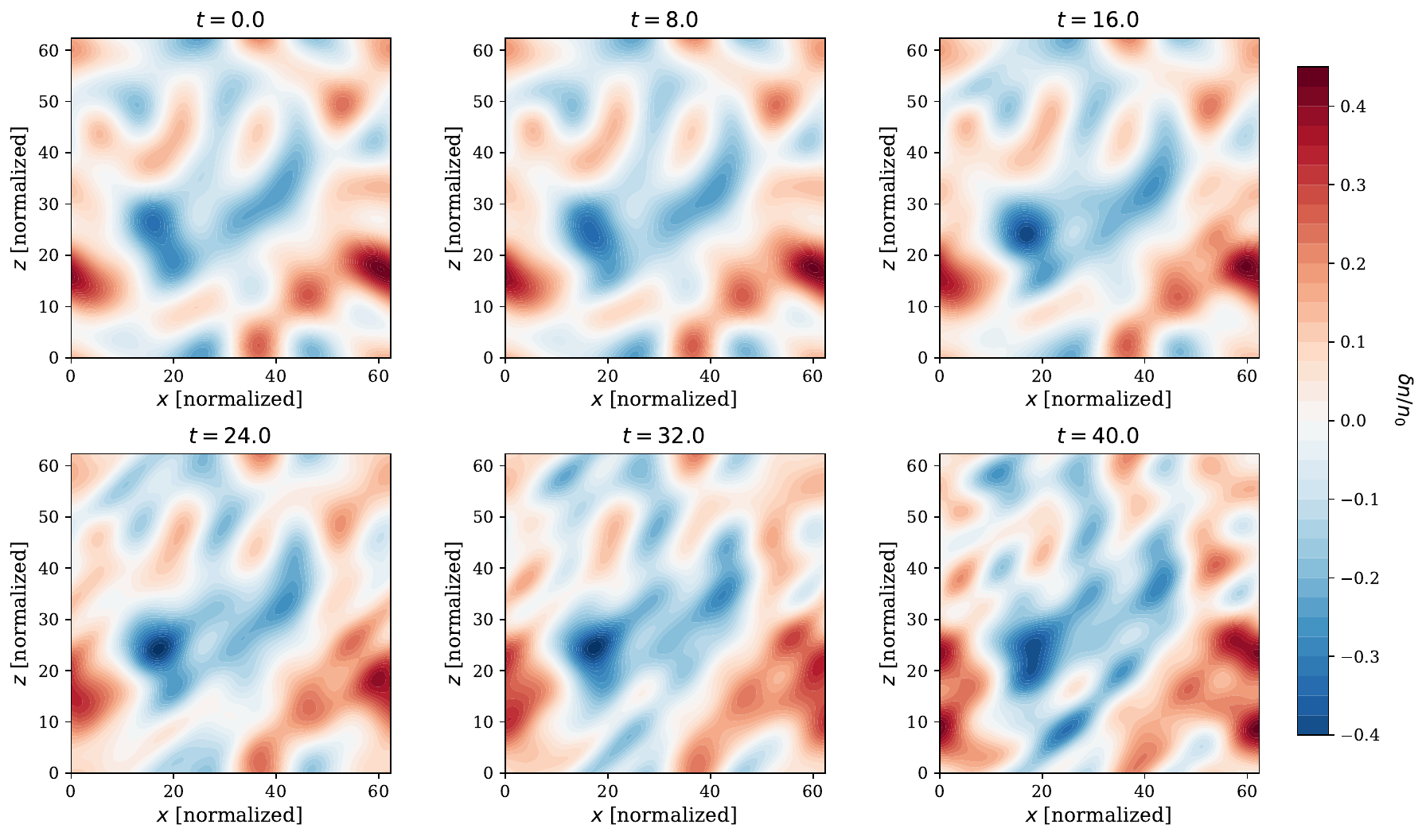}
  \caption{Temporal evolution of the normalized density perturbation $\delta n/n_0$ at
  $t = 0, 8, 16, 24, 32, 40$. Red regions are density enhancements; blue regions are depletions. Persistent cavities (dark blue, $\delta n/n_0\lesssim -0.3$) form co-spatially with magnetic intensity peaks, consistent with ponderomotive expulsion. Amplitudes $|\delta n/n_0|\sim 0.3$--$0.5$ exceed those found in single-wave-packet runs, reflecting the stronger continuous drive from the broadband initialization.}
  \label{fig:density}
\end{figure}

Figure~\ref{fig:density} shows the corresponding evolution of the normalized density perturbation $\delta n/n_0$ at the same six times. The field displays both positive enhancements (red) and negative depletions (blue), with amplitudes reaching $|\delta n/n_0|\sim 0.3$--$0.5$. The spatial pattern closely mirrors, but is not identical to, the magnetic intensity map; in particular, the deepest depletions (dark blue) sit consistently beneath elevated magnetic intensity peaks, the spatial signature of ponderomotive plasma expulsion.

The density structures form on the same timescale as the magnetic ones ($\lesssim 8$ normalized time units) and persist without significant decay to $t = 40$. Prominent depletions co-located with the strongest magnetic filaments are clearly present from $t = 8$ onward and persist through $t = 40$, evolving in shape and depth rather than dispersing. This persistence reflects a sustained quasi-equilibrium between wave pressure and plasma pressure, consistent with the quasi-static limit of equation~(\ref{eq:scaling}).

The amplitude $|\delta n/n_0|\sim 0.3$--$0.5$ is substantially larger than what is typically obtained in single-wave-packet simulations at comparable wave amplitudes \citep{shukla1999}. In a broadband initialization, every KAW mode in the spectrum contributes a gradient of wave pressure at its own scale; these contributions add incoherently to produce a sustained, distributed ponderomotive drive that a single wave packet cannot replicate. The resulting density amplitudes approach the $10$--$30\%$ values reported by MMS in the plasma sheet
\citep{huang2017, gershman2017b}.

\begin{figure}
  \centering
  \includegraphics[width=0.8\textwidth]{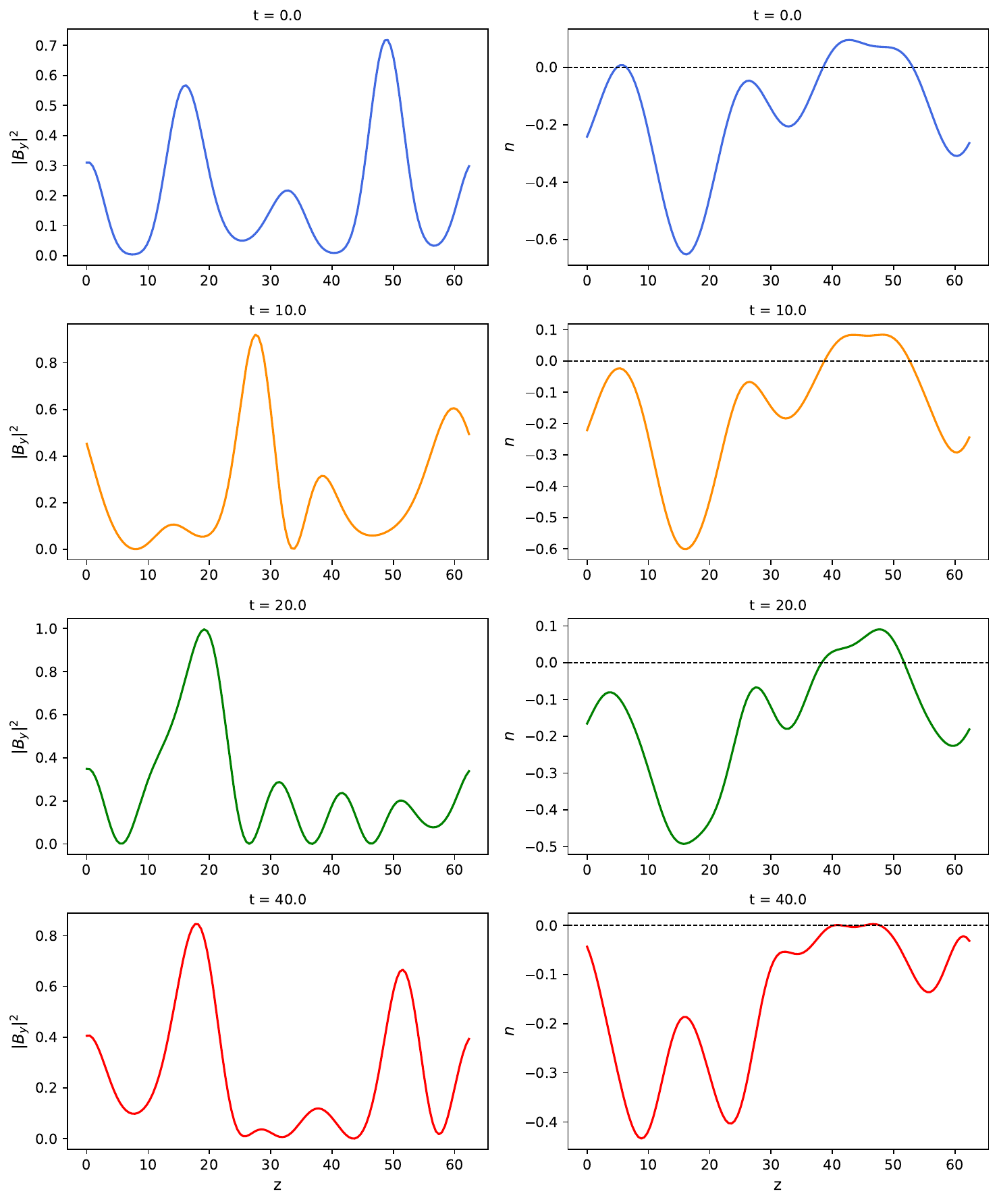}
  \caption{One-dimensional profiles of $|\delta B_y|^2$ (left column) and
  normalized density $n$ (right column) along $z$ at $x = 0$ for $t = 0$, $10$, $20$ and $40$. The density depletions deepen progressively beneath the magnetic intensity peaks, providing direct spatial evidence of ponderomotive plasma expulsion from wave-intense regions.}
  \label{fig:anticorr_slices}
\end{figure}

\begin{figure}
  \centering
 \includegraphics[width=0.8\textwidth]{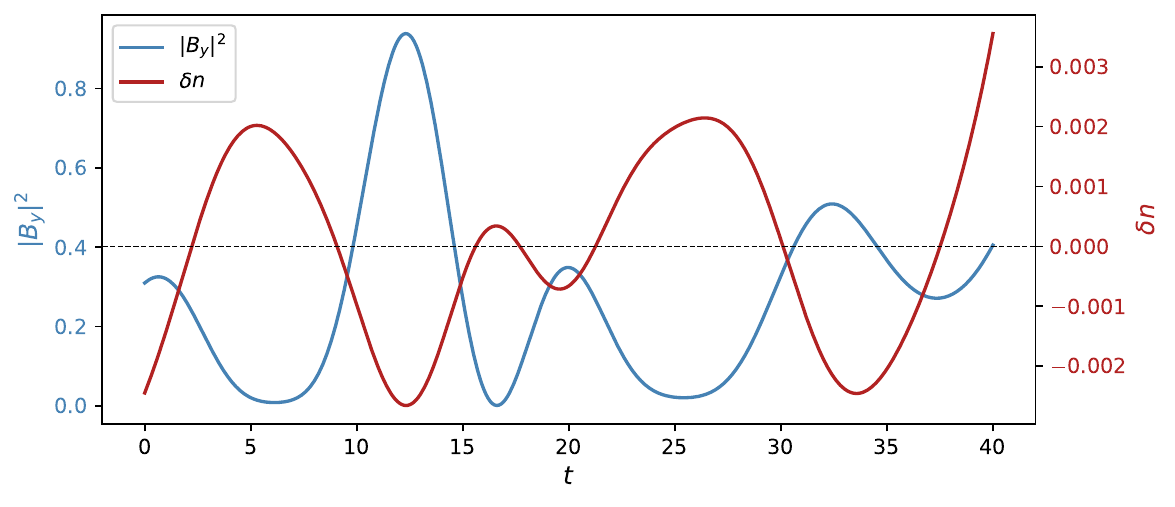}
  \caption{Temporal evolution of $|\delta B_y|^2$ (blue, left axis) and
  $\delta n$ (red, right axis) at a single grid point ($x = 0$, $z = 0$). The two quantities oscillate in approximate anti-phase throughout the simulation, confirming the pointwise ponderomotive anti-correlation predicted by the quasi-static scaling relation (equation~\ref{eq:scaling}).}
  \label{fig:anticorr_temporal}
\end{figure}

To quantify the spatial anti-correlation between wave intensity and density predicted by theory, we compute the Pearson correlation coefficient between $|\delta B_y|^2$ and $\delta n/n_0$ over all grid points in the quasi-steady phase (final two-thirds of the run). The result is $C = -0.022$, which is consistently negative throughout this interval. The small magnitude is expected in a broadband turbulent field: the quasi-static proportionality $n/n_0 \propto -|\psi|^2$ holds mode by mode, but in a multi-mode environment the pixel-by-pixel global average is diluted by cross-mode interference. The spatial co-location of density depletions with wave-intensity peaks visible in Figure~\ref{fig:density} remains the primary diagnostic of ponderomotive expulsion; the negative $C$ confirms the sign of this relationship. Figure~\ref{fig:anticorr_slices} displays one-dimensional profiles of $|\delta B_y|^2$ and $n$ along $z$ at $x = 0$ for $t = 0$, $10$, $20$ and $40$, showing that density depletions deepen beneath intensity peaks as the simulation progresses. Figure~\ref{fig:anticorr_temporal} presents the temporal evolution of $|\delta B_y|^2$ and $\delta n$ at a single grid point ($x = 0$, $z = 0$), where the two quantities oscillate in approximate anti-phase, providing direct pointwise evidence of the ponderomotive coupling predicted by theory.

The ponderomotive force, arising from gradients in the wave intensity, continuously pushes plasma away from regions of high electromagnetic field strength, creating the localized density depletions seen in Figure~\ref{fig:density}. As plasma is expelled outward, the magnetic field becomes concentrated and effectively trapped within these low-density cavities, so the intensity peaks and density depletions are mutually reinforcing structures rather than independent features. This mechanism of ponderomotive cavity formation and magnetic field trapping is well recognized in the context of inertial Alfv\'{e}nic fluctuations in auroral and coronal plasmas, and the present results demonstrate that it operates with equal effect under broadband turbulent driving.

\subsection{Spatial Profiles and Anisotropy}
\label{sec:results:profiles}

\begin{figure}
  \centering
  \includegraphics[width=0.8\textwidth]{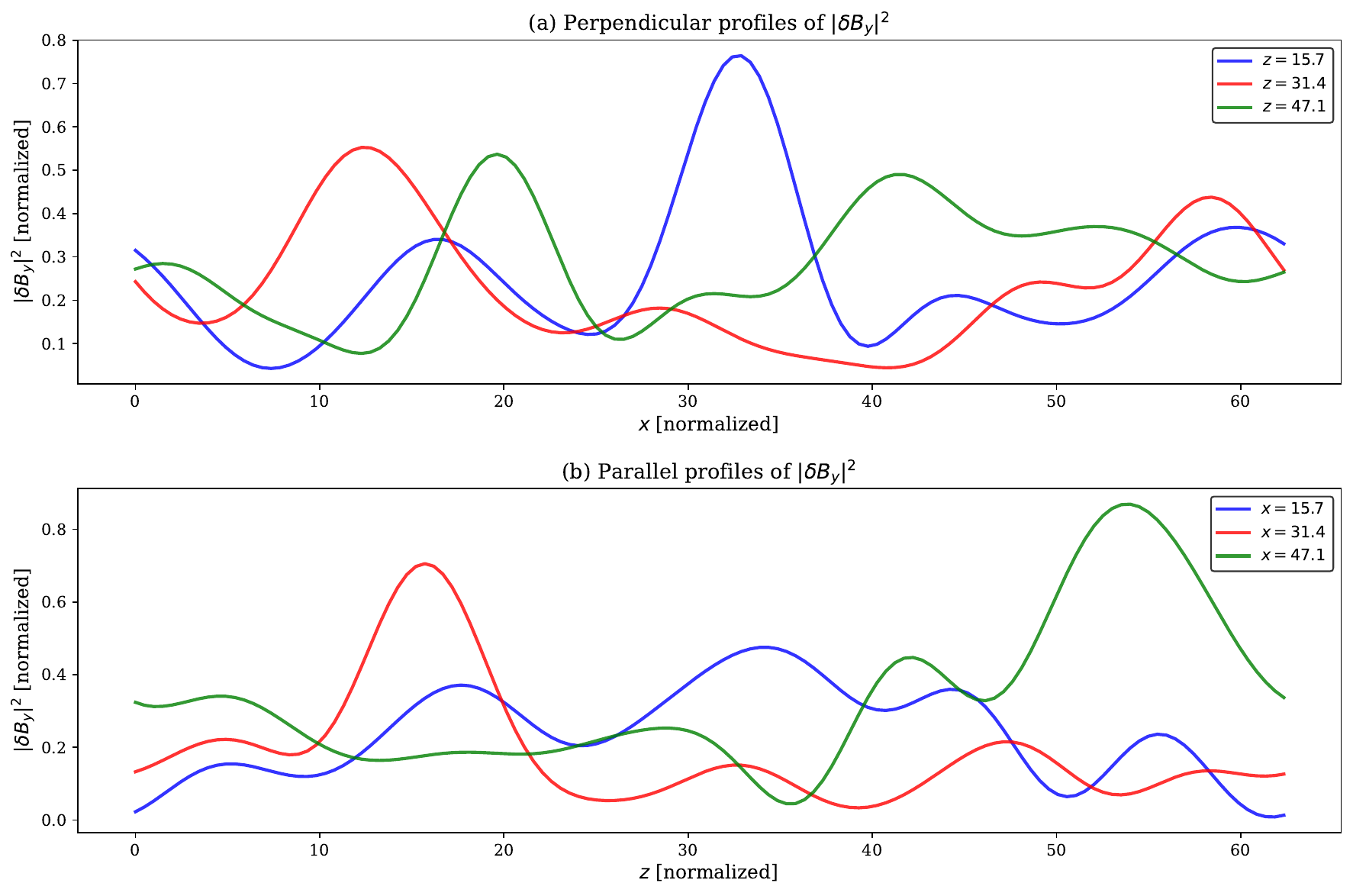}
  \caption{One-dimensional profiles of $|\delta B_y|^2$ (normalized). (a)~Perpendicular cuts
  at $z = 15.7$, $31.4$, and $47.1$. (b)~Parallel cuts at $x = 15.7$, $31.4$, and $47.1$. Peak intensities reach $\sim 1.0$ in both directions with structure widths of $5$--$15$ normalized units, consistent with ion-scale magnetic structures observed by MMS in the plasma sheet boundary layer.}
  \label{fig:profiles}
\end{figure}

Figure~\ref{fig:profiles} shows one-dimensional cuts through $|\delta B_y|^2$ at three cross-sections in the perpendicular direction $x$ (upper panel, fixed $z$) and the parallel direction $z$ (lower panel, fixed $x$). The profiles are irregular rather than periodic or smoothly varying, confirming the turbulent character of the field. Peak intensities vary between cross-sections from near $1.0$ to below $0.4$, providing a direct, quantitative measure of spatial intermittency. The perpendicular cuts show slightly more localized peaks than the parallel cuts, consistent with the anisotropic nature of KAW fluctuations at $k_\perp\rho_i\sim 1$ and with the MNLS dispersion coefficient $P$ (equation~\ref{eq:P}) breaking the $x$--$z$ symmetry at the carrier-wave level \citep{howes2008}.

The characteristic peak widths of $5$--$15$ normalized units correspond to physical scales within the $5$--$20\,d_i$ range of ion-scale magnetic structures observed by multiple MMS studies in the plasma sheet and its boundary layer \citep{huang2017, gershman2017b}.

\subsection{Magnetic Energy Spectra}
\label{sec:results:spectra}

\begin{figure}
  \centering

  \includegraphics[width=0.8\textwidth]{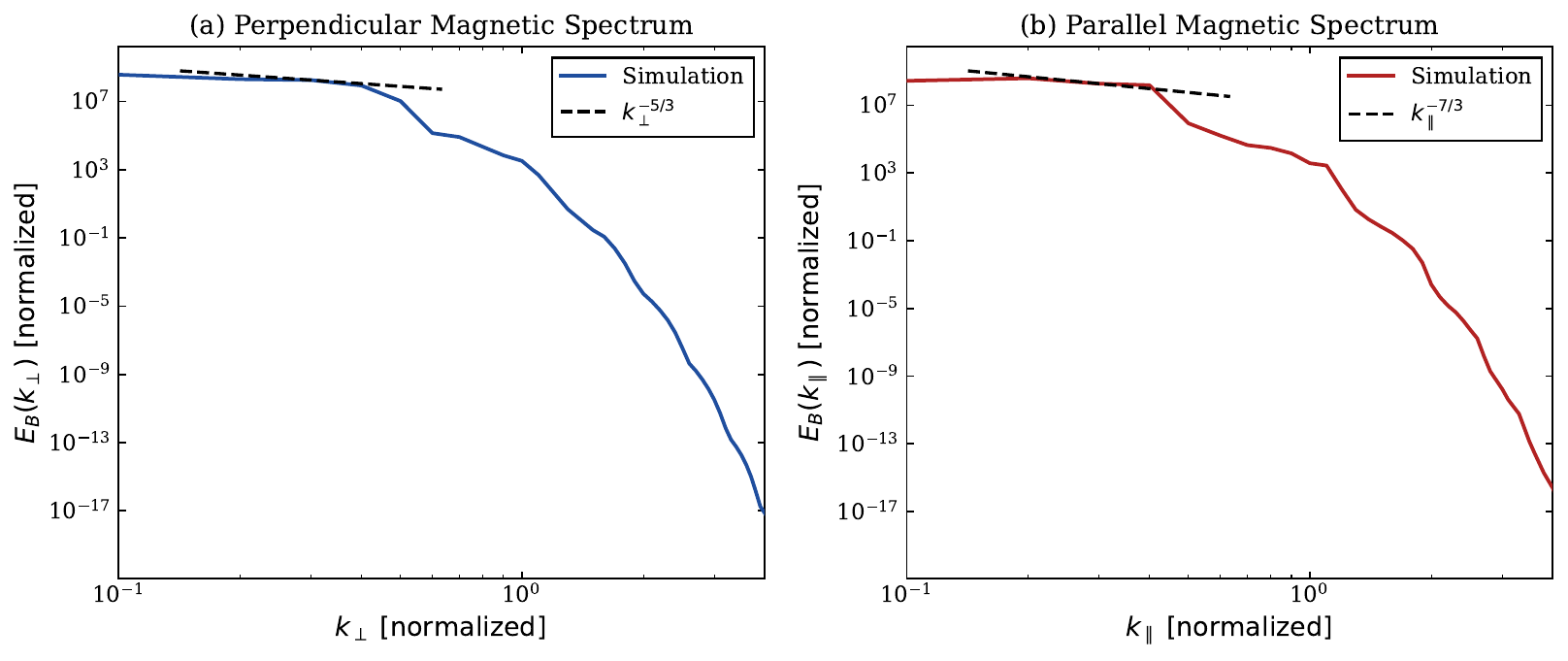}
  \caption{Time-averaged magnetic energy spectra. (a)~Perpendicular spectrum $E_B(k_\perp)$:
  energy is concentrated at large injection scales ($k_\perp \lesssim 0.3$) and falls steeply without forming an extended power-law range before reaching the dissipation floor near $k_\perp \sim 2$. (b)~Parallel spectrum $E_B(k_\parallel)$: qualitatively similar, with a somewhat less steep transition consistent with weaker damping of field-aligned fluctuations. In both panels the dashed line shows the theoretical inertial-range slope ($k^{-5/3}$ perpendicular; $k^{-7/3}$ parallel) anchored at the injection scale as a reference; the immediate and sustained departure of the simulated spectrum from this slope confirms the absence of an inertial range, consistent with the moderate Reynolds number of PSBL turbulence ($\mathrm{R_m}\sim 7$--$110$ from spacecraft observations; see Section~\ref{sec:discussion:Re}).}
  \label{fig:spectra}
\end{figure}

Figure~\ref{fig:spectra} shows time-averaged magnetic energy spectra in the perpendicular ($k_\perp$) and parallel ($k_\parallel$) directions, averaged over the final two-thirds of the run to capture quasi-steady conditions. The Kolmogorov--Goldreich--Sridhar inertial-range slopes $k_\perp^{-5/3}$ and $k_\parallel^{-7/3}$ are shown as dashed reference lines anchored in the injection region; their purpose is to make visible the immediate departure of the simulated spectrum from the theoretical inertial-range prediction, not to assert that such a range is present.

The perpendicular spectrum is approximately flat for $k_\perp < 0.3$, where most energy resides, and steepens sharply beyond $k_\perp\sim 0.3$--$0.5$, dropping by many orders of magnitude before reaching the dissipation floor near $k_\perp\sim 2$. A brief spectral re-enhancement near $k_\perp \approx 0.5$ may reflect a local energy pile-up at the KAW dispersion scale $k_\perp\rho_s\sim 1$; similar pile-ups have been observed in solar wind spectra near the ion gyroscale \citep{bale2005, sahraoui2009, chen2013}. No extended power-law inertial range is present between injection and dissipation.

The parallel spectrum shows a broadly similar shape: energy concentrated at low $k_\parallel$, a broad transition region, and a steep drop at $k_\parallel\gtrsim 0.5$. It falls somewhat less steeply than the perpendicular spectrum at intermediate scales, consistent with weaker damping of field-aligned fluctuations in the absence of the full anisotropic kinetic damping tensor \citep{schekochihin2009}. The absence of an extended inertial range is physically meaningful and is discussed further in Section~\ref{sec:discussion:Re}.

\section{Discussion}
\label{sec:discussion}

\subsection{Ponderomotive Structuring in Broadband Turbulence}
\label{sec:discussion:pondo}

The results demonstrate that ponderomotive density cavity formation is not confined to the idealized case of a single wave packet propagating through a quiescent plasma. In the broadband turbulent simulations presented here, multiple coexisting KAW modes collectively drive a spatially distributed pattern of density depletions and enhancements that emerges spontaneously and persists throughout the run. Every mode in the broadband initialization contributes a gradient of wave pressure at its own scale; these contributions add incoherently, producing a continuous ponderomotive drive spread across the domain that a single wave packet cannot replicate. This accounts for the order-of-magnitude increase in density amplitude relative to single-packet runs at comparable wave amplitudes \citep{shukla1999}, and brings the simulated amplitudes into much closer agreement with the $10$--$30\%$ density fluctuations reported by MMS in the plasma sheet \citep{huang2017, gershman2017b, chettri2025mms}.

\subsection{Reynolds Number Constraints and Spectral Character}
\label{sec:discussion:Re}

The absence of an extended power-law inertial range in the energy spectra (Figure~\ref{fig:spectra}) warrants careful interpretation. The magnetic Reynolds number achievable in a kinetic-scale simulation scales as $R_m \propto N_\mathrm{grid}^{4/3}$ \citep{mininni2004}, placing our $256\times 256$ run in the range $R_m \sim 250$--$370$. This is above the critical threshold for turbulent cascade development in collisionless plasmas ($R_m \approx 107$--$124$;
\citealt{stOnge2020}) but far below the solar wind value of $R_m \sim 10^5$. Spacecraft observations in Earth's plasma sheet yield 
$R_m \sim 7$--$110$ \citep{weygand2007taylor}. Our simulation $R_m \sim 250$--$370$ lies above this observational range; nonetheless, both values belong to the same moderate-$R_m$ regime, orders of magnitude below the solar wind value of $R_m \sim 10^5$--$10^6$. By the Kolmogorov scaling, the ratio of injection to dissipation scales grows as $R_m^{3/4}$, yielding at most one to two decades in $k$-space at these $R_m$ values. Within this narrow window, dissipative mechanisms act on the wave energy almost immediately after injection, and no self-similar inertial range can form. The rapid injection-to-dissipation transition seen in our spectra is therefore not a numerical deficiency but an accurate representation of the physical regime in which PSBL turbulence operates.

In the solar wind, where $R_m \sim 10^5$--$10^6$, the Kolmogorov inertial range spans several decades and a $k^{-5/3}$ spectrum is unambiguous
\citep{bale2005, sahraoui2009}. In the magnetotail, no such separation of scales
exists: the cascade in wavenumber space is compressed to at most one or two decades, and any spectral index extracted from a short power-law segment is strongly sensitive to the fitting range. Our simulations reproduce this compression faithfully, and they caution against over-interpreting spectral slopes derived from short PSBL data intervals. The same constraint applies to the reference slopes shown in Figure~\ref{fig:spectra}: the predictions of \citet{goldreich1995} and
\citet{boldyrev2012} for critically balanced Alfv\'{e}nic turbulence assume
$R_m \gg 1$ and are not directly applicable to the low-$R_m$ PSBL environment.

The perpendicular and parallel magnetic energy spectra in Figure~\ref{fig:spectra} show broadly similar shapes, with no strong preferential enhancement in either direction. The dispersion coefficient $P$ in equation~(\ref{eq:mnls}) breaks $x$--$z$ symmetry at the level of the carrier-wave phase, producing the real-space anisotropy visible in Figure~\ref{fig:profiles}, where perpendicular cuts show slightly more localized peaks than parallel cuts. However, the energy spectrum $E_B(k)\propto|\hat{\psi}(k)|^2$ depends on Fourier amplitudes, not phases, so this phase-level anisotropy does not generate asymmetry between the two power spectra. The deeper spectral anisotropy predicted by Goldreich--Sridhar critical balance ($k_\parallel \propto k_\perp^{2/3}$) is cascade-driven, requiring energy redistribution across many decades of scale by nonlinear shearing. The near-isotropic power spectra seen in Figure~\ref{fig:spectra} are therefore consistent with the isotropic broadband initialization and the sub-Alfv\'{e}nic regime, in which the cascade window is too narrow for such anisotropy to develop.

\subsection{Sub-Alfv\'{e}nic Operation and Astrophysical Relevance}
\label{sec:discussion:astro}

The nonlinearity parameter $\chi_\mathrm{NL}\approx 0.25$ throughout places the system in the sub-Alfv\'{e}nic (or weak turbulence) regime, where wave periods are shorter than the nonlinear interaction time and energy transfer proceeds primarily through resonant three-wave interactions
\citep{galtier2000}. Ponderomotive density structures form efficiently here because the wave
coherence time is long enough to sustain a quasi-static pressure gradient, in agreement with the analytical predictions of \citet{shukla1999}.

The broader implications extend to several astrophysical environments. In the solar corona and solar wind, KAW ponderomotive structuring is a candidate mechanism for density striations that produce enhanced radio scintillation \citep{cranmer2012, matthaeus2011}. In the inner heliosphere, coupled KAW and ion-acoustic wave dynamics have been identified in Parker Solar Probe observations \citep{chettri2024raa}, lending further support to the ubiquity of ponderomotive coupling across heliospheric environments. In the interstellar medium, kinetic-scale Alfv\'{e}nic turbulence governs pulsar scattering and cosmic-ray transport through pitch-angle scattering \citep{lazarian2006, yan2004}. In galaxy clusters, kinetic instabilities akin to KAW ponderomotive coupling are invoked to explain density fluctuations below the Coulomb mean free path in the hot intracluster medium \citep{brunetti2007, schekochihin2009}. The ICM parameters ($\beta\sim 50$--$100$; length scales of kiloparsecs) differ enormously from the PSBL studied here, but the governing kinetic-scale wave equations share the same mathematical structure. The central conclusion of the present work, that broadband KAW turbulence spontaneously organizes into spatially coherent density structures through ponderomotive forcing and sustains that organization over many wave periods, therefore applies to these environments by appropriate parameter scaling \citep{verscharen2019, schekochihin2009}.

\subsection{Limitations}
\label{sec:discussion:limits}

Several limitations of the present model motivate follow-on work. The simulations are two-dimensional, so three-dimensional effects such as field-aligned focusing of wave energy and anisotropic cascade development \citep{goldreich1995} are absent. The parallel-propagation approximation adopted for the density equation~(\ref{eq:msw}) neglects cross-field acoustic transport; at higher $\beta$, where $C_s$ approaches $v_A$, the full Laplacian $\nabla^2$ should be retained. The MNLS framework (equation~\ref{eq:mnls}) incorporates Landau damping through a scalar coefficient $\Gamma$; it does not resolve the full anisotropic, wave-vector-dependent kinetic damping tensor captured by gyrokinetic descriptions
\citep{schekochihin2009, howes2008}. Ion temperature is held constant, suppressing the pressure
anisotropy feedback and mirror-mode-like saturation that could develop in a fully kinetic treatment at higher $\beta$. Wave-particle resonances, particularly the electron Landau channel identified by \citet{chen2019} and the ion heating inferred from MMS energy-flux measurements
\citep{gershman2017}, are not self-consistently resolved by the fluid-envelope approach.

\section{Summary and Conclusions}
\label{sec:conclusions}

We have presented two-dimensional pseudospectral simulations of the MNLS--magnetosonic system under plasma-sheet-boundary-layer conditions ($\beta\sim 0.1$--$0.3$), initialized with a broadband turbulent spectrum spanning many simultaneously active wave modes. The governing equations follow the modified nonlinear Schr\"{o}dinger framework of \citet{chettri2025damped}, and the key questions concern how turbulent multi-mode driving, rather than a single wave packet, shapes density structuring, spectral character, and Reynolds-number physics under low-$\beta$ PSBL conditions. The main findings are as follows.

\begin{enumerate}

\item Total energy is conserved to within $0.085\%$ over the full integration to $t = 40$
normalized units, with hyperviscous dissipation confined to grid scales, confirming the numerical reliability of the RK4 pseudospectral solver; in the damped run, an additional $\sim 4\%$ of the magnetic energy is lost over the same interval, consistent with the modest role of collisionless dissipation at these parameters.

\item The simulation operates in the sub-Alfv\'{e}nic turbulent regime throughout
($\chi_\mathrm{NL}\approx 0.25$, well below the strong-turbulence threshold), placing the results in the range relevant to quieter PSBL intervals and to sub-Alfv\'{e}nic turbulence in a variety of astrophysical environments.

\item Spatially intermittent, filamentary magnetic structures emerge from the broadband
initialization within a few wave periods and persist throughout the run. Their perpendicular scales, $\sim 10$--$20\,\rho_s$, are consistent with ion-scale magnetic structures observed by MMS in the plasma sheet boundary layer \citep{huang2017, gershman2017b, chettri2025mms}.

\item Persistent density cavities with amplitudes $|\delta n/n_0|\sim 0.3$--$0.5$ form
co-spatially with magnetic intensity peaks, confirming ponderomotive expulsion of plasma from wave-intense regions. These amplitudes are an order of magnitude larger than those produced by single-wave-packet simulations at comparable wave amplitudes, demonstrating that broadband turbulent driving substantially enhances ponderomotive density structuring.

\item The magnetic energy spectra exhibit a rapid injection-to-dissipation transition without
an extended inertial range, consistent with the moderate magnetic Reynolds number of both the simulation ($\mathrm{R_m}\sim 250$--$370$) and the actual PSBL plasma ($\mathrm{R_m}\sim 7$--$110$ from spacecraft observations). The absence of a power-law inertial range is an intrinsic feature of this turbulent regime, not a numerical artefact.

\end{enumerate}

These results show that ponderomotive density structuring by KAW turbulence is not restricted to the idealized single-wave-packet limit but operates equally well under broadband turbulent driving. The spectral character of kinetic-scale turbulence in low-Reynolds-number environments such as the PSBL is qualitatively distinct from that of the solar wind, and spectral slopes extracted from short data intervals in this regime should be interpreted with care. The MNLS pseudospectral approach used here is fast enough to permit systematic parameter surveys of how density structuring depends on $\beta$, wave amplitude, and injection scale.

Future work should extend the simulations to three dimensions to capture field-aligned focusing, incorporate kinetic physics through hybrid or gyrokinetic approaches to resolve Landau damping more completely, and carry out direct event-by-event comparisons with MMS burst-mode intervals in which KAW activity and density structuring are simultaneously resolved.

\section*{Acknowledgements}

Authors MKC, VS, and BS acknowledge the University Grants Commission (UGC), India, for support through the Non-NET Fellowship programme. One of the authors RM acknowledges support from IUCAA Pune through visiting associate program.

\printcredits

\bibliographystyle{cas-model2-names}
\bibliography{refs}

@article{bale2005,
  author  = {Bale, S. D. and Kellogg, P. J. and Mozer, F. S.
             and Horbury, T. S. and Reme, H.},
  title   = {Measurement of the electric fluctuation spectrum of
             magnetohydrodynamic turbulence},
  journal = {Physical Review Letters},
  year    = {2005},
  volume  = {94},
  pages   = {215002},
  doi     = {10.1103/PhysRevLett.94.215002}
}

@article{boldyrev2012,
  author  = {Boldyrev, S. and Perez, J. C.},
  title   = {Spectrum of kinetic-{Alfv\'{e}n} turbulence},
  journal = {The Astrophysical Journal Letters},
  year    = {2012},
  volume  = {758},
  pages   = {L44},
  doi     = {10.1088/2041-8205/758/2/L44}
}

@article{chettri2024raa,
  title={Nonlinear Coupling of Kinetic {Alfv\'{e}n} Waves and Ion Acoustic Waves in the inner Heliosphere},
  author={Chettri, Mani K and Shrivastav, Vivek and Mukherjee, Rupak and Gaur, Nidhi and Sharma, RP and Singh, Hemam D},
  journal={Research in Astronomy and Astrophysics},
  volume={24},
  number={10},
  pages={105009},
  year={2024},
  publisher={National Astronomical Observatories, CAS and IOP Publishing}
}

@article{brunetti2007,
  author  = {Brunetti, G. and Lazarian, A.},
  title   = {Compressible turbulence in galaxy clusters: physics
             and stochastic particle re-acceleration},
  journal = {Monthly Notices of the Royal Astronomical Society},
  year    = {2007},
  volume  = {378},
  pages   = {245--275},
  doi     = {10.1111/j.1365-2966.2007.11771.x}
}

@article{chaston2008,
  author  = {Chaston, C. C. and Bonnell, J. W. and Carlson, C. W.
             and McFadden, J. P. and Mozer, F. and Ergun, R. E.
             and Strangeway, R. J.},
  title   = {Kinetic effects in the acceleration of auroral electrons
             in small-scale {Alfv\'{e}n} waves: a {FAST} case study},
  journal = {Geophysical Research Letters},
  year    = {2008},
  volume  = {35},
  pages   = {L02105},
  doi     = {10.1029/2007GL032603}
}

@article{chen2013,
  author  = {Chen, C. H. K. and Boldyrev, S. and Xia, Q.
             and Perez, J. C.},
  title   = {Nature of sub-proton scale turbulence in the solar wind},
  journal = {Physical Review Letters},
  year    = {2013},
  volume  = {110},
  pages   = {225002},
  doi     = {10.1103/PhysRevLett.110.225002}
}

@article{chen2019,
  author  = {Chen, C. H. K. and Klein, K. G. and Howes, G. G.},
  title   = {Evidence for electron {Landau} damping in space plasma
             turbulence},
  journal = {Nature Communications},
  year    = {2019},
  volume  = {10},
  pages   = {740},
  doi     = {10.1038/s41467-019-08435-3}
}

@article{chettri2025damped,
  title={Damped Kinetic {Alfv\'{e}n} Waves in Earth's Magnetosheath: Numerical Simulations and MMS Observations},
  author={Chettri, Mani K and Singh, Hemam D and Shrivastav, Vivek and Singh, Britan and Mukherjee, Rupak},
  journal={arXiv preprint arXiv:2512.09828},
  year={2025}
}

@article{chettri2025mms,
  title={MMS Observations of Kinetic {Alfv\'{e}n} Wave Turbulence and Steep Kinetic-Range Spectra in the Outer Plasma Sheet Boundary Layer},
  author={Chettri, Mani K and Singh, Hemam D and Mukherjee, Rupak},
  journal={arXiv preprint arXiv:2603.07969},
  year={2026}
}

@article{cranmer2012,
  author  = {Cranmer, S. R. and {van Ballegooijen}, A. A.},
  title   = {Proton, electron, and ion heating in the fast solar
             wind from nonlinear coupling between {Alfv\'{e}nic}
             and fast-mode turbulence},
  journal = {The Astrophysical Journal},
  year    = {2012},
  volume  = {754},
  pages   = {92},
  doi     = {10.1088/0004-637X/754/2/92}
}

@article{denton2010,
  author  = {Denton, R. E. and others},
  title   = {Multiple harmonic {ULF} waves in the plasma sheet
             boundary layer: instability analysis},
  journal = {Journal of Geophysical Research},
  year    = {2010},
  volume  = {115},
  pages   = {A12224},
  doi     = {10.1029/2010JA015505}
}

@article{du2011,
  author  = {Du, A. M. and others},
  title   = {Fast tailward flows in the plasma sheet boundary layer
             during a substorm on 9 {March} 2008: {THEMIS}
             observations},
  journal = {Journal of Geophysical Research},
  year    = {2011},
  volume  = {116},
  pages   = {A03216},
  doi     = {10.1029/2010JA015969}
}

@article{galtier2000,
  author  = {Galtier, S. and Nazarenko, S. V. and Newell, A. C.
             and Pouquet, A.},
  title   = {A weak turbulence theory for incompressible
             magnetohydrodynamics},
  journal = {Journal of Plasma Physics},
  year    = {2000},
  volume  = {63},
  pages   = {447--488},
  doi     = {10.1017/S0022377899008284}
}

@article{gershman2017,
  author  = {Gershman, D. J. and others},
  title   = {Wave-particle energy exchange directly observed in a
             kinetic {Alfv\'{e}n}-branch wave},
  journal = {Nature Communications},
  year    = {2017},
  volume  = {8},
  pages   = {14719},
  doi     = {10.1038/ncomms14719}
}

@article{gershman2017b,
  author  = {Gershman, D. J. and others},
  title   = {Magnetic holes in a kinetic turbulent plasma},
  journal = {Journal of Geophysical Research: Space Physics},
  year    = {2017},
  volume  = {122},
  pages   = {11108--11120},
  doi     = {10.1002/2017JA024263}
}

@article{goldreich1995,
  author  = {Goldreich, P. and Sridhar, S.},
  title   = {Toward a theory of interstellar turbulence.
             {II.}~Strong {Alfv\'{e}nic} turbulence},
  journal = {The Astrophysical Journal},
  year    = {1995},
  volume  = {438},
  pages   = {763--775},
  doi     = {10.1086/175121}
}

@article{hasegawa1976parametric,
  author  = {Hasegawa, A. and Chen, L.},
  title   = {Parametric decay of ``kinetic {Alfv\'{e}n} wave'' and
             its application to plasma heating},
  journal = {Physical Review Letters},
  year    = {1976},
  volume  = {36},
  pages   = {1362--1365},
  doi     = {10.1103/PhysRevLett.36.1362}
}

@article{howes2008,
  author  = {Howes, G. G. and Cowley, S. C. and Dorland, W.
             and Hammett, G. W. and Quataert, E.
             and Schekochihin, A. A.},
  title   = {A model of turbulence in magnetized plasmas:
             implications for the dissipation range in the solar
             wind},
  journal = {Journal of Geophysical Research: Space Physics},
  year    = {2008},
  volume  = {113},
  pages   = {A05103},
  doi     = {10.1029/2007JA012665}
}

@article{huang2017,
  author  = {Huang, S. Y. and others},
  title   = {{MMS} observations of ion-scale magnetic holes in the
             plasma sheet},
  journal = {Geophysical Research Letters},
  year    = {2017},
  volume  = {44},
  pages   = {7131--7139},
  doi     = {10.1002/2017GL074778}
}

@article{lazarian2006,
  author  = {Lazarian, A. and Beresnyak, A.},
  title   = {Cosmic ray scattering in compressible turbulence},
  journal = {Monthly Notices of the Royal Astronomical Society},
  year    = {2006},
  volume  = {373},
  pages   = {1195--1202},
  doi     = {10.1111/j.1365-2966.2006.11093.x}
}

@article{leamon1999,
  author  = {Leamon, R. J. and Smith, C. W. and Ness, N. F.
             and Wong, H. K.},
  title   = {Dissipation range dynamics: kinetic {Alfv\'{e}n} waves
             and the importance of $\beta_e$},
  journal = {Journal of Geophysical Research},
  year    = {1999},
  volume  = {104},
  pages   = {22331--22344},
  doi     = {10.1029/1999JA900158}
}

@article{matthaeus2011,
  author  = {Matthaeus, W. H. and Velli, M.},
  title   = {Who needs turbulence? {A} review of turbulence effects
             in the heliosphere and on the fundamental process of
             reconnection},
  journal = {Space Science Reviews},
  year    = {2011},
  volume  = {160},
  pages   = {145--168},
  doi     = {10.1007/s11214-011-9793-9}
}

@article{mininni2004,
  author  = {Mininni, P. D. and G\'{o}mez, D. O.
             and Mahajan, S. M.},
  title   = {Dynamo action in magnetohydrodynamics and
             {Hall}-magnetohydrodynamics},
  journal = {The Astrophysical Journal},
  year    = {2004},
  volume  = {587},
  pages   = {472--481},
  doi     = {10.1086/368181}
}

@article{roberts2018,
  author  = {Roberts, O. W. and others},
  title   = {Observation of an {MHD} {Alfv\'{e}n} vortex in the
             slow solar wind},
  journal = {Journal of Geophysical Research: Space Physics},
  year    = {2018},
  volume  = {123},
  pages   = {6998--7009},
  doi     = {10.1029/2018JA025248}
}

@article{sahraoui2009,
  author  = {Sahraoui, F. and Goldstein, M. L. and Robert, P.
             and Khotyaintsev, Yu. V.},
  title   = {Evidence of a cascade and dissipation of solar-wind
             turbulence at the electron gyroscale},
  journal = {Physical Review Letters},
  year    = {2009},
  volume  = {102},
  pages   = {231102},
  doi     = {10.1103/PhysRevLett.102.231102}
}

@article{schekochihin2009,
  author  = {Schekochihin, A. A. and others},
  title   = {Astrophysical gyrokinetics: kinetic and fluid turbulent
             cascades in magnetized weakly collisional plasmas},
  journal = {The Astrophysical Journal Supplement Series},
  year    = {2009},
  volume  = {182},
  pages   = {310--377},
  doi     = {10.1088/0067-0049/182/1/310}
}

@inproceedings{shrivastava2015,
  author    = {Shrivastava, G. and Shrivastava, J. and Ahirwar, G.},
  title     = {Kinetic {Alfv\'{e}n} wave in the presence of kappa
               distribution function in the plasma sheet boundary
               layer},
  booktitle = {AIP Conference Proceedings},
  year      = {2015},
  volume    = {1670},
  pages     = {030031},
  doi       = {10.1063/1.4926722}
}

@article{shukla1999,
  author  = {Shukla, P. K. and Stenflo, L.},
  title   = {Nonlinear propagation of kinetic {Alfv\'{e}n} waves},
  journal = {Physics of Plasmas},
  year    = {1999},
  volume  = {6},
  pages   = {4120--4124},
  doi     = {10.1063/1.873674}
}

@article{stawarz2017,
  author  = {Stawarz, J. E. and others},
  title   = {Magnetospheric {Multiscale} analysis of intense
             field-aligned {Poynting} flux near the {Earth}'s
             plasma sheet boundary},
  journal = {Geophysical Research Letters},
  year    = {2017},
  volume  = {44},
  pages   = {7106--7113},
  doi     = {10.1002/2017GL074109}
}

@article{stOnge2020,
  author  = {{St-Onge}, D. A. and Kunz, M. W. and Squire, J.
             and Schekochihin, A. A.},
  title   = {Fluctuation dynamo in a collisionless, weakly
             magnetized plasma},
  journal = {Journal of Plasma Physics},
  year    = {2020},
  volume  = {86},
  pages   = {905860503},
  doi     = {10.1017/S0022377820000860}
}

@article{terradas2022,
  author  = {Terradas, J. and Soler, R. and Oliver, R.
             and Ballester, J. L.},
  title   = {Parametric decay of standing {Alfv\'{e}n} waves in a
             coronal loop},
  journal = {Astronomy \& Astrophysics},
  year    = {2022},
  volume  = {660},
  pages   = {A136},
  doi     = {10.1051/0004-6361/202142975}
}

@article{verscharen2019,
  author  = {Verscharen, D. and Klein, K. G. and Maruca, B. A.},
  title   = {The multi-scale nature of the solar wind},
  journal = {Living Reviews in Solar Physics},
  year    = {2019},
  volume  = {16},
  pages   = {5},
  doi     = {10.1007/s41116-019-0021-0}
}

@misc{verscharen2024,
  author       = {Verscharen, D. and others},
  title        = {Parametric decay instability of circularly
                  polarised {Alfv\'{e}n} waves with a spectrum of
                  wavevectors: {3D} hybrid simulations},
  year         = {2024},
  howpublished = {arXiv preprint arXiv:2403.08179},
  doi          = {10.48550/arXiv.2403.08179}
}

@article{weygand2007taylor,
  title={Taylor scale and effective magnetic Reynolds number determination from plasma sheet and solar wind magnetic field fluctuations},
  author={Weygand, James M and Matthaeus, WH and Dasso, S and Kivelson, MG and Walker, RJ},
  journal={Journal of Geophysical Research: Space Physics},
  volume={112},
  number={A10},
  year={2007},
  publisher={Wiley Online Library}
}

@article{wygant2002,
  author  = {Wygant, J. R. and others},
  title   = {Evidence for kinetic {Alfv\'{e}n} waves and parallel
             electron energization at 4--6~$R_E$ altitudes in the
             plasma sheet boundary layer},
  journal = {Journal of Geophysical Research: Space Physics},
  year    = {2002},
  volume  = {107},
  pages   = {1201},
  doi     = {10.1029/2001JA900113}
}

@article{yan2004,
  author  = {Yan, H. and Lazarian, A.},
  title   = {Cosmic-ray scattering and streaming in compressible
             magnetohydrodynamic turbulence},
  journal = {The Astrophysical Journal},
  year    = {2004},
  volume  = {614},
  pages   = {757--769},
  doi     = {10.1086/423733}
}

@article{zhou2012,
  author  = {Zhou, X.-Z. and others},
  title   = {Emergence of the active magnetotail plasma sheet
             boundary from transient, localized ion acceleration},
  journal = {Journal of Geophysical Research},
  year    = {2012},
  volume  = {117},
  pages   = {A10216},
  doi     = {10.1029/2012JA018171}
}

\end{document}